# Multimodal Tip-Enhanced Nonlinear Optical Nano-Imaging of Plasmonic Silver Nanocubes

*Chih-Feng Wang and Patrick Z. El-Khoury[*]*

Physical Sciences Division, Pacific Northwest National Laboratory, P.O. Box 999, Richland, WA 99352

[*]patrick.elkhoury@pnnl.gov (PZE)

**ABSTRACT**. Optical field localization at plasmonic tip-sample nanojunctions has enabled high spatial resolution chemical analysis through tip-enhanced linear optical spectroscopies, including Raman scattering and photoluminescence. Here, we illustrate that nonlinear optical processes, including parametric four-wave mixing (4WM), second harmonic/sum-frequency generation (SHG and SFG), and two-photon photoluminescence (TPPL), can be enhanced at plasmonic junctions and spatio-spectrally resolved simultaneously with few-nm spatial resolution under ambient conditions. More importantly, through a detailed analysis of our spectral nano-images, we find that the efficiencies of the local nonlinear signals are determined by sharp tip-sample junction resonances that vary over the few-nanometer length scale because of the corrugated nature of the probe. Namely, plasmon resonances centered at or around the different nonlinear signals are tracked through TPPL, and they are found to selectively enhance nonlinear signals with closely matched optical resonances.

**TOC Graphic**

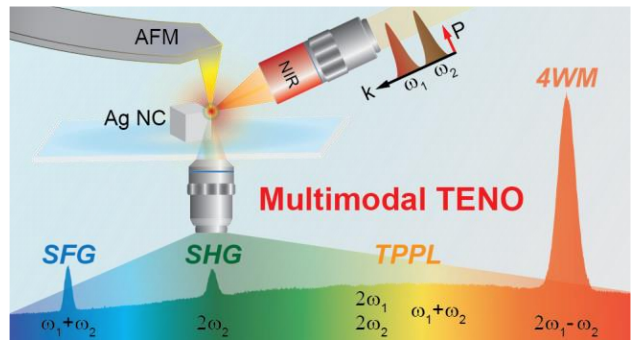

Continuous wave and pulsed laser irradiation of metallic nanostructures can lead to optical field nano-localization and enhancement. The latter has been exploited in plasmon-enhanced optical spectroscopy and microscopy to achieve single molecule detection sensitivity and sub-nm spatial resolution in chemical imaging.[1] Ultra-high spatial resolution and equivalently ultrasensitive nano-imaging and nano-spectroscopy is nowadays routinely achieved through the combination of optical spectroscopy, e.g., Raman and photoluminescence, and scanning probe microscopy. Indeed, tip-enhanced photoluminescence (TEPL[2]) and tip-enhanced Raman spectroscopy (TERS[3]) have found applications in fields as diverse as quantum information science and technology, biological imaging, and heterogeneous catalysis.[4-6] Tip-enhanced nonlinear optical measurements are more scarce in comparison,[6] even though proof of principle measurements have been reported many years ago.[6-10]

Spatially localized and enhanced nonlinear optical signals has been previously observed in tip-enhanced nonlinear nanoscopy.[11] Very recently, we illustrated that tip-enhanced 4-wave mixing[6, 12] can be used to visualize local optical fields with sub-2 nm spatial resolution under ambient conditions.[13] In the prior measurements, we found that the recorded near field images trace local optical fields defined by the interaction between the plasmonic probe and nanoparticles (cubes) on the substrate.[13] In other words, the optical response at junction plasmons formed between the tip (both its apex or shaft) and the underlying plasmonic nanostructure dictate the intensity patterns observed in the nonlinear nano-images. A similar conclusion was reached when (linear) molecular Raman scattering was used to image plasmonic nanostructures,[14] including silver nanocubes and faceted nanoparticles.[15] Overall, since the nano-structural makeup of the tip inevitably dictates the observables,[16] some broadening of the edge response (at regions in space where the shaft of the tip interacts with the edge of the nanoparticle) is expected for tips with larger diameters compared to high aspect ratio tips.[17] That said, how corrugations at the apex of a plasmonic probe affect junction plasmon resonances sampled at different tip positions atop a plasmonic nanoparticle (such as the case in this study) remains unclear. Both the effect of the tip on the recorded nano-images as well as local junction plasmon resonances that vary over the nanometer length scale are important to keep in mind in going through the ensuing discussion.

Here, we revisit tip-enhanced nonlinear optical nano-imaging and nano-spectroscopy. We use a familiar platform, namely crystalline plasmonic silver nanocubes, to closely inspect and better-understand the correlations between various nonlinear signals, i.e., parametric four-wave mixing (4WM), second harmonic/sum-frequency generation (SHG and SFG), and two-photon photoluminescence (TPPL), on the nanometer scale. Analysis of our spectrally-resolved images allows us to conclude that sharp localized plasmon resonances that vary over the few-nanometer length scale dictate the relative efficiencies of the different spatio-spectrally distinct nonlinear optical signals. These variations in local plasmon resonances arise from the corrugated nature of the plasmonic probe used in this study (see the supporting information section).

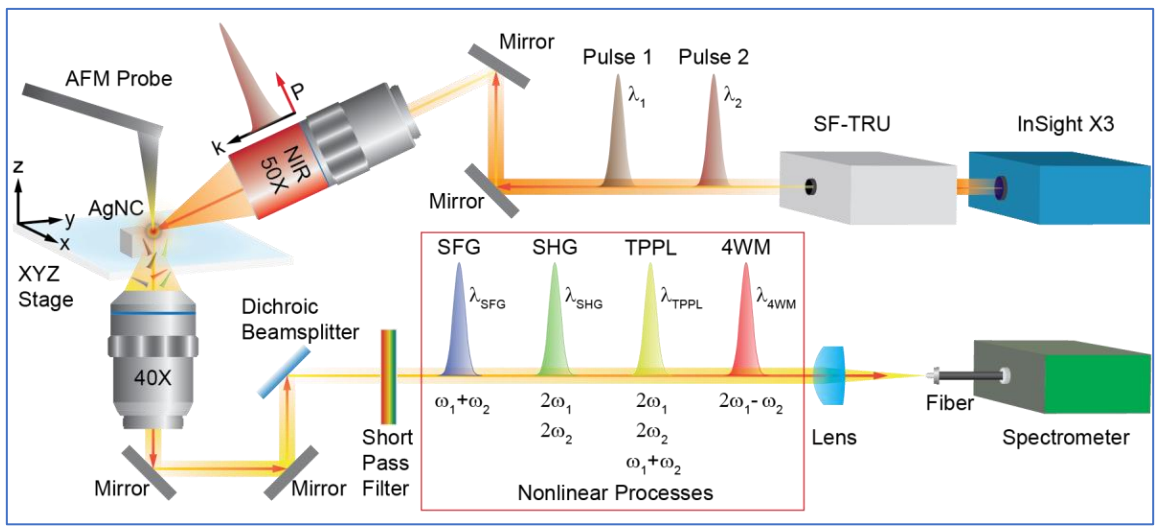

**Figure 1.** Schematic representation of our nonlinear nanoscopy setup. The light source consists of a commercially available femtosecond light source that outputs 2 difference laser beams, herein centered at 1045 and 820 nm. The two collinear/time-overlapped pulses are polarized along the long axis of the tip, and focused onto its apex using a 50X air objective. The scattered radiation is collected using a bottom optical axis, filtered off a dichroic beamsplitter and through a short pass filter, prior to focusing the optical signals into a fiber of a fiber-coupled detection system. More details are given in the main text. AgNC: silver nanocube; SFG: sum-frequency generation; SHG: second harmonic generation; TPPL: two-photon photoluminescence; 4WM: four-wave mixing.

Figure 1 shows a schematic representation of our nonlinear nanoscopy setup. A commercially available laser system (InSight X3, Spectra Physics) serves as a light source and outputs two laser beams at a repetition rate of 80 MHz. The first is fixed at 1045 nm with a pulse width of ~150 fs at the sample position, which is achieved using a built-in chirp pre-compensation module. The second output is tunable in the 680-1300 nm spectral region with a ~100-120 fs pulse duration at the sample. The two beams are rendered collinear and are time-overlapped using a commercially available spectral focusing/time recombination unit (SF-TRU, Newport). The two collinear beams ($\lambda_1$ = 1045 nm; $\lambda_2$ = 820 nm) are subsequently focused onto the apex of a silver-coated AFM tip (100 nm thick coating) using a near-IR objective (Mitutoyo, 50X, NA=0.42) at an angle of ~65° with respect to the surface normal. The polarization states of the incident beams are set along the long axis of the plasmonic probe. The scattered nonlinear optical response is collected using a separate (bottom) objective (Olympus, 40X, NA=0.95). A combination of a long pass dichroic beam splitter and a short pass filter is used to reject the incident near-IR lasers, prior to focusing the nascent signals into a fiber-coupled spectrometer (Andor, Shamrock SR-300i) coupled to an EM-CCD (Andor, Newton DU970P-BVF). Tip-enhanced optical signals were recorded using an intermittent contact mode feedback, as described in prior work from our group.[13, 18] In this scheme, signals were recorded both when (i) the tip is in contact with the substrate, and (ii) the tip is 20-40 nm away from the sample at the same position. The latter (far-field) response is subtracted from the former that contains both near- and far-field contributions to allow us to inspect pure near-field images. The supporting information section emphasizes the aforementioned protocol, all while highlighting additional tip-enhanced multimodal nonlinear optical images recorded using different tips and analyzing different plasmonic nanocubes.

Figure 2 shows a topographic AFM image of a single silver nanocube (a) and simultaneously recorded near-field SFG (b), SHG (c), and 4WM (d) images. The 100 nm silver nanocubes we use (SCPH100-1M from nanoComposix) are highly uniform, as specified by the manufacturer and shown elsewhere.[13, 15] As such, deviations from ideal cubical shape in both the topographic as well as the near field images are related to the nano-structural makeup of the probe (see supporting information section). Tip convolution in topographic AFM mapping is well-understood. That said, its analogue in tip-enhanced optical spectroscopy is less thoroughly investigated.[17] Furthermore, fine nanometeric corrugations are known to sharpen the spatial resolution in tip-enhanced optical spectroscopy,[16] but there is no systematic approach that may be used to map local resonances over the relevant single/few nanometer length scale. Here, the spatially broadened signal in the immediate vicinity of the cube (near the 4 edges that are observable in the nonlinear images in Figures 2) arises from the interaction between the shaft of the plasmonic probe with the edges of the nanoparticle. Images recorded with finer probes[13, 15] (also see the supporting information section), as well as the sharp drop in nonlinear optical signals when the tip traverses from the edges to atop the nano-cube corroborate this assignment. We stress that dim signals observed when the tip is atop the nanocube is not caused by the excitation/collection scheme used herein (side/bottom; see Figure 1). Rather, the images recorded trace the spatially varying local optical field magnitudes under our experimental conditions. As illustrated elsewhere,[13-15] TERS images of plasmonic silver nanocubes that were collected using side excitation-back scattered collection also reveal that the edges of the nanocubes are selectively enhanced. Note that in the absence of the plasmonic probe, the enhanced local optical field of the cube itself is only selectively enhanced at its edges, as shown in the supporting information section.

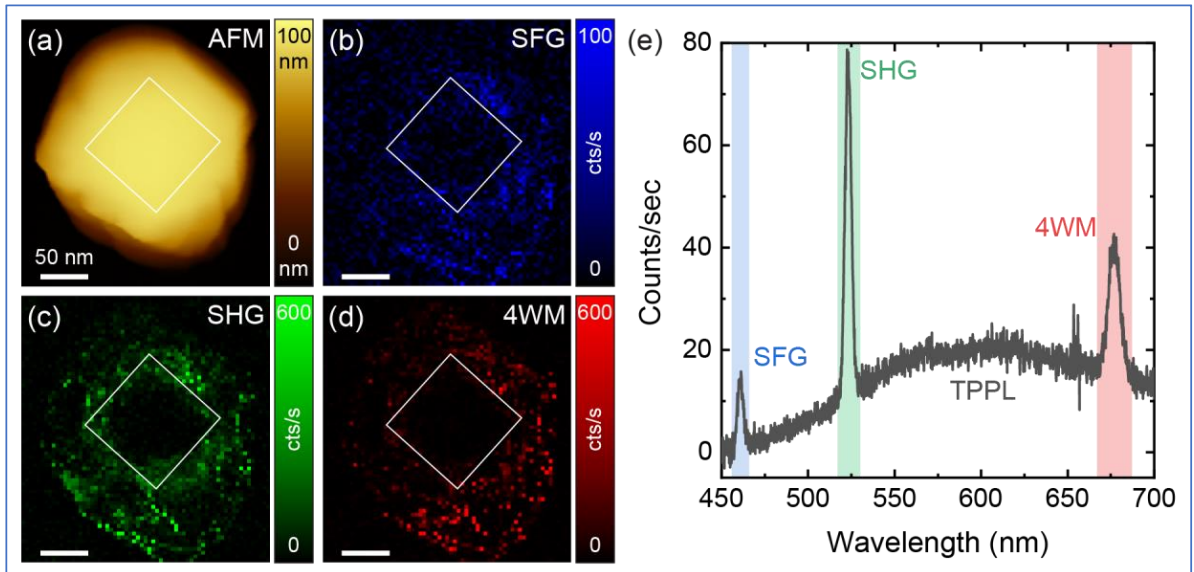

**Figure 2.** Atomic force microscopy (a) and near-field SFG (b), SHG (c), as well as 4WM (d) images of a 100 nm silver nanocube. The SFG, SHG, and 4WM images were spectrally integrated in the 456-466 nm, 517-530 nm, and 667-687 nm regions, respectively. The latter is indicated using color-coded rectangular boxes in panel e, which also shows a spatially-integrated spectrum. Note that the squares in panels a-d approximately mark the 4 edges of the cube prior to convolution (see text for more details). Conditions: the incident pulses $\lambda_1/\lambda_2$ are centered at 820 nm/1045 nm with average powers of 0.25 mW per arm. The signals were time-integrated for 0.1 s at every pixel and the lateral step size (both horizontal and vertical) used was 4 nm.

Prior to a more detailed inspection of the recorded spectra that vary over the few-nm length scale, we note that the different nonlinear signals that can be clearly distinguished in the spatially averaged spectrum (see Figure 2e) ride on a broad background that arises from tip-enhanced TPPL. Much like its linear analogue (PL),[19] we will show that this signal tracks the local plasmon resonance, herein with few nanometer spatial resolution. To the best of our knowledge, no other approach may be used to track tip-sample junction plasmons with the herein demonstrated spatial resolution. Given the nano-corrugated nature of the sputtered probe, as gauged from the recorded topographic and nonlinear optical signals (Figure 2), different junction plasmon resonances are operative at different positions of the tip near/atop the nanocube. Even though the latter has been invoked in the past to explain observables in TERS,[14,20] direct measurements of the local plasmonic resonance without molecular reporters that can alter the response have not been possible, until now.

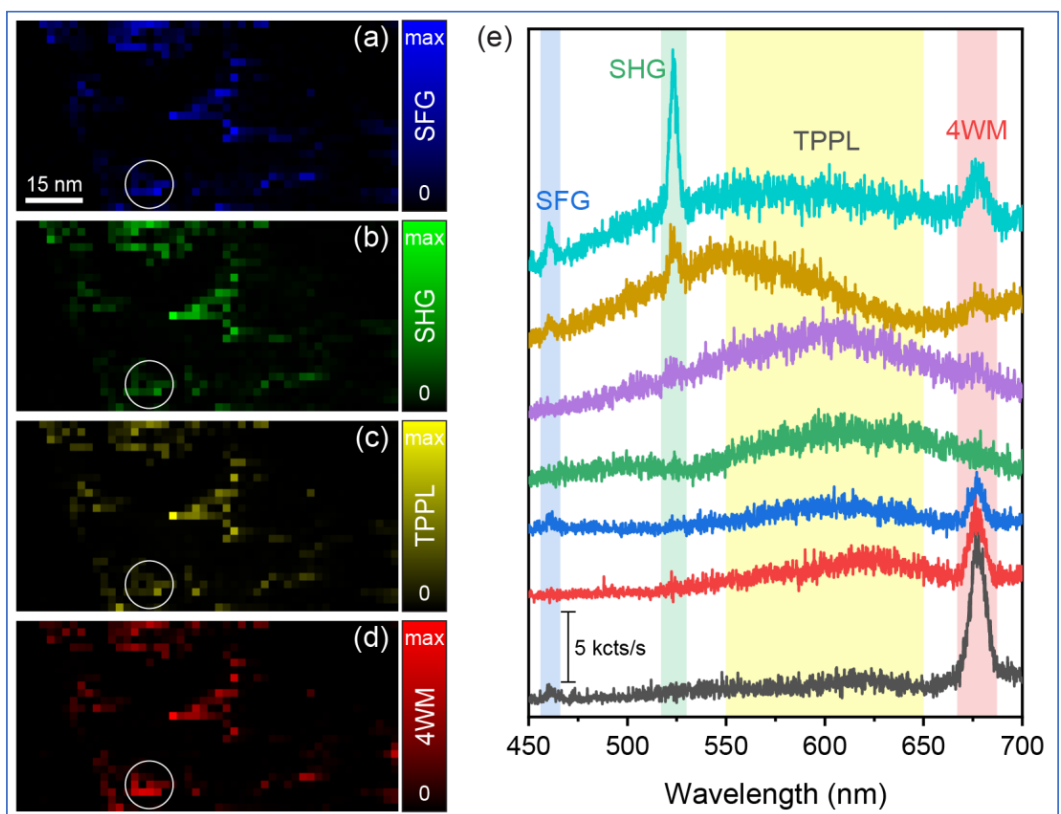

**Figure 3.** Near-field SFG (a), SHG (b), TPPL (c), and 4WM (d) images recorded around the edge of a 100 nm silver nanocube. The SFG, SHG, TPPL, and 4WM images were spectrally integrated in the 456-466 nm, 517-530 nm, 550-650, and 667-687 nm regions, respectively. The latter is indicated using color-coded rectangular boxes in panel e, which also shows selected single pixel spectra recorded at different spatial positions. One region is highlighted using a white circular mark and tracked across panels a-d (see text for more details). Conditions: the incident pulses $\lambda_1/\lambda_2$ are centered at 820 nm/1045 nm with average powers of 0.25 mW per arm. The signals were time-integrated for 0.1 s at every pixel and the later step size (both horizontal and vertical) used was 2 nm.

A high spatial resolution multimodal nonlinear optical image recorded around the edge of a nanocube is shown in Figure 3. The pixelated response we observe indicates a pixel-limited lateral spatial resolution (< 2 nm) in these measurements, which is consistent with recent observations.[13] Comparison of the simultaneously recorded SHG, SFG, 4WM, and TPPL signals also reveals that some areas (e.g., within the white circular region in a-d) feature distinct relative intensities at the resonance energies that correspond to the different nonlinear optical processes. This is bolstered in Figure 3e, where selected single pixel spectra taken at different spatial positions (see Figure S3c) are shown on the same plot. Another interesting observation in Figure 3e has to do with the broad two-photon photoluminescence signals, the resonance maxima of which are distinct at different regions in space. Based on (i) prior correlated single particle optical dark field micro-scattering-PL measurements,[19] (ii) the observation that the two-photo photoluminescence image traces the tip-sample nano-junction in space (see Figure 3c and Figure S2h), and (iii) the spatially varying resonance maxima and pixelated response that can be traced back to the nano-structural makeup of the probe (see Figure 3e), we can assign the spatially varying tip-enhanced TPPL signal to the spatially varying tip-sample nanojunction plasmon resonance. More in support of this assignment follows. Overall, it is important to concede that the nano-structural makeup of the probe may evolve over the time scale of our measurement because of the typical powers used and also as a result of the (albeit intermittent) contact mode AFM feedback. That said, the images shown in Figure 3 are simultaneously recorded. As such the physical insights gained by our measurements are not affected by the spatiotemporally varying junction resonances.

The recorded spatio-spectrally distinguishable nonlinear signals are expected to be selectively enhanced by energy matched junction plasmons, herein tracked through the TPPL signal. In other words, spatially varying resonance matching conditions must be operative: a plasmon peaking near one of the three nonlinear signals should optimally enhance the nonlinear process that most closely matches its energy. One way of examining the latter without bias/hand selecting spectra to make a case is through 2D correlation analysis. A representative 2D cross-correlation map is shown in Figure 4a. In this plot, the diagonal correlation ($\rho_{j,k} = 1$), SFG, SHG, and 4WM signals are all visible. Any positive off-diagonal correlation between the background (TPPL/junction plasmon) signal on one hand and various nonlinear signals on the other indicates that the two signals simultaneously increase in intensity. Cross-correlation slices taken at different wavelengths corresponding to the 4WM, SHG, and SFG resonances are informative in this regard. The cuts shown in Figure 4b reveal a strong correlation between the three different nonlinear optical signals and relatively sharp tip-substrate plasmon resonances (here appearing as backgrounds) that peak in their immediate vicinities. Note that the dip in one signal at cuts taken at another, e.g. in the 4WM signal at cuts taken at the SFG/SHG resonances do not indicate a negative correlation, but rather a non-correlated response, given the magnitude of the dips. From this analysis, the different nonlinear signals appear to be indeed enhanced by plasmons with closely matched energies.

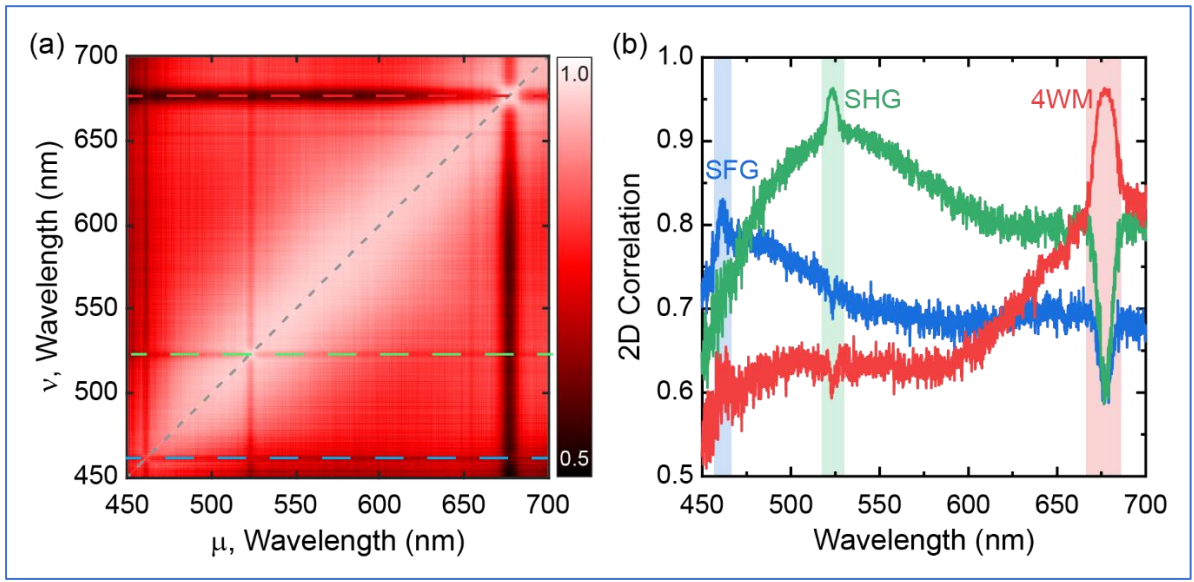

**Figure 4.** 2D cross-correlation map ($\rho_{j,k} = \sigma_{jk}^2/\sigma_{jj}\cdot\sigma_{kk}$ in a) and slices (b). The cuts shown in panel b were spectrally integrated (in a 3 nm region). The red, green, and blue dash lines are taken at 4WM, SHG, and SFG signal maxima, respectively. See text for more details.

Several aspects of the measurements reported herein bolster the difference between microscopic and nanoscopic selection rules in nonlinear optical microscopy and spectroscopy. First, the near field nonlinear signals are scattered; they are not collected in the back or forward directions, but rather from the bottom following side excitation (see Figure 1). Classically, no signal is measurable in this geometry, as shown in the supporting information measurement for the cube itself. Second, the images traced using the different nonlinear signals are identical (see Figure 2), notwithstanding the different selection rules of the second-order *vs* third order signals. This is because all the recorded nonlinear signals are optimally enhanced at plasmonic tip-nanoparticles junctions, where classical selection rules are lifted. In effect, the recorded maps track the junctions formed between the tip (its apex and shaft) and the four different sides of the nanocube. In the absence of the probe, a dipolar response would be expected from the cube itself (see supporting information). Differences in the relative efficiencies of the SFG, SHG, and 4WM signals are otherwise dictated by sharp resonances that vary in space because of the difference in the atomic makeup of the nanojunctions at difference tip/landing positions.

In summary, we describe multimodal tip-enhanced nonlinear optical microscopy measurements that track different nonlinear signals (4WM, SFG, SHG, and TPPL) at plasmonic nanojunctions defined by a nano-corrugated probe and a crystalline silver nanocube. Analysis of the recoded spectral images reveals that the TPPL signals, which exhibit substantial spatio-spectral variations, track spatially varying junction plasmons. These plasmons, the energies of which are distinct when different regions of the plasmonic probe interact with the underlying silver nanocube, selectively resonantly enhance local nonlinear optical signals that exhibit closely matched resonance maxima. The latter is confirmed through 2D correlation analysis of the recorded spectral nano-images. Beyond the scope of this work, spatially varying resonances may account for the observations of tip position dependent backgrounds in TERS spectra (both Stokes and anti-Stokes regions). More generally, the interpretation of spectral (and background) fluctuations in TERS/TEPL spectra and images have to account for the possibility of spatially varying plasmon resonances. In the future, it would be interesting to understand the correlations between nanoscale chemical images (TERS) with local resonance maps (TE-TPPL). On a final note, it is important to recognize that three different resonances contribute to the recorded nanoscale linear/nonlinear signals: (i) the tip resonance, (ii) the nanocube resonance, and (iii) the tip-nanocube junction plasmon resonance that herein varies in space. Under the experimental conditions used here, the junction plasmon governs the observables. Using different tips and nanoparticles inevitably changes the picture, and in some cases may be used to selectively track the local optical fields of the particle itself. These are important considerations to keep in mind in designing tip enhanced linear and nonlinear optical measurements aimed at tracking the optical fields of plasmonic metal nanoparticles.

## METHODS

Silver nanocube samples were prepared by drop-casting a 20 μL solution of cubes (SCPH100-1M from nanoComposix) on glass coverlips (TedPella, 260150). After the colloidal solution was air-dried, the substrate was rinsed with excess amounts of ethanol.

Nanoscale-resolved multimodal nonlinear images were recorded using a previously described AFM-Raman optical system.[21-22] The general experimental approach and schematic are described in the main text and the parameters are given in figure captions. The tips we used consist of silicon probes (Nanosensors, ATEC-NC) sputtered with Ag (100 nm thickness). Typical oscillation amplitudes in our AFM measurements were on the order of 20-40 nm using probes with resonance frequencies of ~300 kHz.

## SUPPORTING INFORMATION

Far field signals recorded from a plasmonic silver probe and from an isolated silver nanocube under the same conditions used throughout this work and additional images that were recorded using both an imperfect and a sharp probe.

## ACKNOWLEDGMENTS

The authors acknowledges support from the United States Department of Energy, Office of Science, Office of Basic Energy Sciences, Division of Chemical Sciences, Geosciences & Biosciences. Some of the instrumentation that was required to do the measurements described in this work was purchased and developed through funding from the Laboratory Directed Research and Development (LDRD) program at Pacific Northwest National Laboratory.

## AUTHOR INFORMATION

**Corresponding Author**

*patrick.elkhoury@pnnl.gov

The authors declare no competing financial interest.